\documentclass[%
 reprint,
 amsmath,amssymb,
 aps,
floatfix,
]{revtex4-2}

\usepackage{graphicx}
\usepackage{dcolumn}
\usepackage{bm}
\usepackage{xcolor}
\usepackage{soul}

\begin{document}

\preprint{APS/123-QED}

\title{A Computational Study of Cluster Dynamics in Structural Lubricity: Role of Cluster Rotation}

\author{Wai H. Oo}%
\author{Mehmet Z. Baykara}
\affiliation{%
Department of Mechanical Engineering, University of California at Merced, Merced, California 95343, USA
}%

\author{Hongyu Gao}
\email{Corresponding author: hongyu.gao@uni-saarland.de}
\affiliation{%
Department of Materials Science and Engineering, Saarland University, Saarbrücken, Saarland 66123, Germany
}%


%


\date{\today}

\begin{abstract}
We present a computational study of sliding between gold clusters and a highly oriented pyrolytic graphite substrate, a material system that exhibits ultra-low friction due to structural lubricity.
By means of molecular dynamics, it is found that clusters may undergo spontaneous rotations during manipulation as a result of elastic instability, leading to attenuated friction due to enhanced interfacial incommensurability.
In the case of a free cluster, shear stresses exhibit a non-monotonic dependency on the strength of the tip-cluster interaction, whereby rigid clusters experience nearly constant shear stresses.
Finally, it is shown that the suppression of the translational degrees of freedom of a cluster's outermost-layer can partially annihilate out-of-plane phonon vibrations, which leads to a reduction of energy dissipation that is in compliance with Stokesian damping. It is projected that the physical insight attained by the study presented here will result in enhanced control and interpretation of manipulation experiments at structurally lubric contacts.
\end{abstract}


\maketitle


\section{Introduction}
Superlubricity describes a state of ultra-low friction between two surfaces under relative motion \cite{Baykara2018APR}, a particularly desirable property in mechanical systems for reduction of energy consumption \cite{Holmberg2017F}.
To distinguish the superlubricity that arises due to structural incommensurability between contacting surfaces (rather than due to a lubricant film or an externally applied electric/magnetic field) \cite{Shinjo1993SS}, the term  `structural lubricity' was introduced \cite{Mueser2004EL}.
The physics underlying structural lubricity involves a systematic annihilation of mean lateral forces at the interface due to geometric incommensurability during sliding motion, which manifests in exceptionally low interaction energy corrugations.
A key resulting characteristic of structural lubricity is that friction scales sub-linearly with contact area \cite{Mueser2001PRL,Dietzel2013PRL}, i.e., $F\propto{A}^\gamma$, where exponent $\gamma$=0.5 serves as an upper-bound, representing a scenario where one of the contacting surfaces is amorphous.
Although it has been generally accepted to be very challenging to implement in practical applications, promising progresses has been achieved in terms of preserving the structurally lubric state at macro-scale contacts \cite{Liu2012PRL,Berman2015S} and under ambient conditions \cite{Cihan2016NC,Deng2018N}.

A particular limitation of structural lubricity arises at extended contact sizes. Specifically, when the elastic correlation length associated with the slider is exceeded, interfacial structural defects in the form of dislocations cause the sliding cluster to `pin', leading to a breakdown of the ultra-low friction state \cite{Sharp2016PRB,Monti2020ACSNano}. In such cases, shear stresses will no longer depend on interfacial alignment but saturate at a finite value instead, namely the Peierls stress \cite{Sharp2016PRB}.
Even at small length scales where clusters are nominally rigid, pinning can still take place at the edges of a cluster, in the absence of pinning agents such as adsorbed layers \cite{Mueser2001PRL} and unsaturated chemical bonds \cite{Dietzel2017ACSNano}.
Not surprisingly, lateral forces will rise at the contact lines where the stress gradients are the highest, particularly so when those contact lines are normal to the sliding direction.
As a recent proof of this argument, Gao and M\"user \cite{Gao2022FC} demonstrated that one possible origin of instabilities can be from quasi-discontinuous dynamics of the moir\'e patterns (manifesting atom normal displacements) near a propagating contact line, during the sliding of a semi-infinite gold slab on a graphite substrate.
Such sliding discontinuity can be attributed to the symmetry breaking nature of a multi-layer gold cluster with its (111) surface parallel to the graphite basal plane.
Consequently, the mean friction forces exhibit only weak dependence on sliding speed, or namely Coulomb-type friction \cite{Prandtl1928ZAMM,Tomlinson1929}.

It should be noted that variations of mean kinetic friction beyond statistical errors are routinely observed in nanoscale experiments \cite{Dietzel2008PRL, Dietzel2018N}, for which various origins can exist, e.g. contact aging \cite{Feldmann2014PRL} and varying cluster morphology \cite{Wijn2012PRB}.
Additionally, at a structurally well-defined interface free of contaminants, it is well-known that friction can vary as a function of sliding orientation as well as misfit angles \cite{Dienwiebel2004PRL, Wijn2012PRB}.
Finally, along a given path, a cluster can rotate (i)  in response to an external magnetic field \cite{Cao2022PRX} or (ii) in the presence of a driving atomic force microscopy (AFM) tip, due to weakly constrained in-plane rotational degree of freedom (RDoF), provided that point-contact is formed between the driving tip and the cluster's surface \cite{Filippov2008PRL}.
Here comes a critical question: Does cluster rotation always result in a lowering of the interfacial lattice mismatch and thus lead to an elevated energy barrier, which would manifest as increased friction \cite{Filippov2008PRL}?
Despite the critical nature of this question, discussions about in-plane rotational friction and its coupling with translational friction at nanoscale contacts are limited, mainly because the interplay is uneasy to clarify experimentally or to be properly addressed by means of simplified theoretical models \cite{Prandtl1928ZAMM,Frenkel1939}.

Motivated by the question above, we present here molecular dynamics (MD) simulations of nanoscale cluster manipulation performed on the representative gold-graphite system, which has been shown to exhibit structural lubricity  under both ultra-high vacuum (UHV) and ambient conditions \cite{Wijn2012PRB, Dietzel2013PRL, Cihan2016NC,Gao2022FC}.
In particular, kinetic friction has been studied in the presence of spontaneous cluster rotations, whereby the tip-cluster interaction has been modeled in the form of a harmonic spring linked directly to the cluster's topmost-layer.
Additionally, the effect of boundary control of atom translational degrees of freedom on interfacial phonon vibrations (and thereby the modeled shear stresses) has been analyzed via phonon spectra converted from Fourier-transformed instantaneous forces.
Overall, we expect this study to provide physical insights into friction control at structurally lubric nanoscale contacts.

\section{Methodology}

\begin{figure}[ht]
\centering
\includegraphics[width=0.45\textwidth]{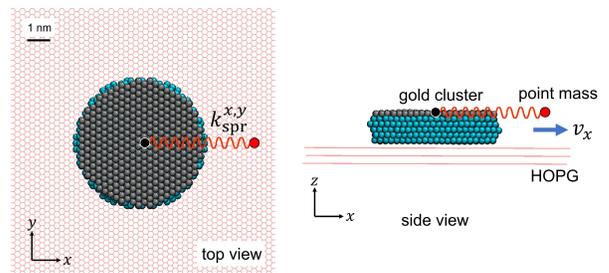}
\caption{Snapshots of the MD model at rest. During sliding, the point mass drags the cluster at a constant speed of 20 m/s along the $x$-axis. The spring constants, both in the lateral ($x$) and the transverse ($y$) directions, vary in the range of 1.6-1600 N/m.}
\label{fig_snapshot}
\end{figure}

The MD simulations model the dynamic sliding process of a circular gold cluster on a highly oriented pyrolytic graphite (HOPG) substrate in a contaminant-free environment.
As the model in Fig.~\ref{fig_snapshot} shows, the two surfaces forming the gold-graphite interface are structurally well-defined.
The cluster moves along the $x$-axis, dragged by a point mass via a spring that connects to the center-of-mass (COM) of its topmost-layer.
The stiffness of the driving spring ($k_{\rm spr}$) along both the $x$- and $y$-axis is the same, varying in a range of 1.6-1600 N/m.
The gold cluster (with a radius of 2.85 nm) consists of six atomic layers in the (111) configuration stacked parallel to the graphite basal plane.
The three-layer graphite substrate has in-plane dimensions of 11.5×11.6 nm$^2$, by means of which periodic boundary conditions are applied.
The atomic interactions for isolated gold and graphite are described by an EAM potential \cite{Zhou2004PRB} and the AIREBO potential \cite{Stuart2000JCP}, respectively.
The interaction between the two materials is described by a Morse potential \cite{de-la-Rosa-Abad2016RSC}. 
The point mass does not directly interact with any atoms in the system except by means of the spring attached to the top surface of the gold cluster.

The COM of the bottommost-layer of graphite is fixed throughout the simulations, whereas the related atoms can still move freely according to Newton's equation of motion.
The temperature of the system is maintained at 0.1 K through a Langevin thermostat applied to the mid-graphite layer in only the $z$-direction. 
Such a cryogenic environmental control aims only to get rid of the thermal noise so as to achieve an improved signal-to-noise ratio in the simulated values.
The topmost-layer of the gold cluster is set to be either rigid or free:
A `rigid' layer signifies that the relative displacements of the associated atoms are completely suppressed such that the whole layer acts as a unit; while, a `free' layer, together with the rest of the free atoms in the system, indicates that no constraints have been put on the motion of the atoms.
In both scenarios, the RDoFs of the cluster are not constrained.
The point mass slides along the armchair direction of the graphite substrate, i.e., the $x$-axis, at a constant speed of 20 m/s as default, unless specified otherwise.
The total sliding distance is over 500 nm, of which data points in the steady-state (typically within the latter 400 nm of the total sliding distance) are used in the calculation of mean values.
In the normal direction, the cluster and the graphite substrate interact only through adhesion without any external loads applied.
The simulation timestep is 1.0 fs.
All MD simulations are carried out using the open-source code LAMMPS \cite{Plimpton1995JCP}.

The calculation of shear stresses ($\tau$) is conducted according to the definition: ${\tau} = {\langle} F {\rangle} / A$, where $F$ denotes the friction force, $A$ the contact area, and $\langle...\rangle$ a time average.
Interfacial forces acting between the cluster and substrate are derived directly from the interfacial interaction with statistical errors ($\Delta O^2$) obtained from an integration of a time auto-correlation function
$C_{OO}(t)=\langle O(t')O(t'+t)\rangle$ via $\Delta O^2=\frac{2}{\Gamma}\int_0^\Gamma C_{OO}(t){\rm d}t$, where $O$ denotes an observable and $\Gamma$ the simulation time.
The mean lateral forces calculated as above are confirmed to be nearly identical to the mean elastic restoring forces on the driving spring, as well as the mean resisting forces from the bottommost-layer of the graphite substrate, thus proving mechanical consistency.
Moreover, mean lateral forces are comparable to those converted from the dissipated energy adsorbed by the thermal reservoir via $P = \langle F \rangle \cdot v$, where $P$ is the dissipated friction power that can be calculated via $P = \gamma \sum_{i \in {\rm mid}} (m \langle v_{iz}^2 \rangle - k_{\rm B}T)$.
In the latter expression, $\gamma$ is the damping coefficient and $v_{iz}$ is the thermal velocity of atom $i$ in the $z$-direction.
The contact area is a summation of per-atom areas of gold atoms in direct contact with graphite, which is estimated in terms of an average nearest-neighbor distance ($\sim$2.8 \AA).
Within the elastic limit, the contact areas vary only marginally between different simulation runs.

\section{Results and discussion}
\subsection{Role of driving spring stiffness}

\begin{figure}[ht]
\centering
\includegraphics[width=0.4\textwidth]{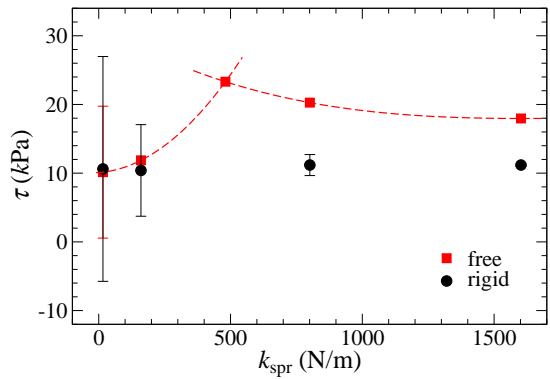}~~~~~~
\caption{Shear stress of a gold cluster sliding on graphite as a function of spring constant $k_{\rm spr}$. The cluster's topmost layer is either free (red squares) or constrained (black circles) to a rigid, yet freely rotating plane. The driving spring moves along the armchair direction of graphite at a constant speed of 20 m/s. Error bars smaller than the signal size are not shown, same as in the following figures.}
\label{fig_sig_k}
\end{figure}

When the driving spring stiffness $k_{\rm spr}$ is much stronger than the interfacial interaction between the cluster and the substrate, the latter dominates the effective contact stiffness ($k_{\rm eff}$) assuming that (i) the interfacial interaction can also be represented by a virtual spring and (ii) the two springs are connected in series \cite{Carpick1997APL}.
The model spring constants in the simulations span a wide range, with the lower bound in the proximity of typical experimental values \cite{Dong2013JASTA, Socoliuc2004PRL}, i.e., $\mathcal{O}(10~{\rm N/m})$.
As shown in Fig.~\ref{fig_sig_k}, for the case of a rigid topmost cluster layer,  MD simulations reveal that the mean shear stresses are independent of the driving spring stiffness.
Within the framework of single-asperity contact, this result is consistent with what is predicted by the Prandtl model in the regime $\eta \leq$ 1 ($\eta=4\pi^2E_0/k_{\rm eff}a^2$), where the overdamped point mass slides smoothly with ultra-low energy dissipation \cite{Socoliuc2004PRL}.
Please note that detailed calculations based on the Prandtl model can be found in the Appendix.
In terms of this model, further increasing $k_{\rm spr}$ should have a marginal effect on the mean shear stresses; whereas further decreasing $k_{\rm spr}$ may cause mean shear stresses to rise at some point due to enhanced elastic instability, provided that the corrugation potential landscape of the substrate remains unchanged.

Unless the cluster grows epitaxially from the substrate such that the interfacial structural mismatch is fairly small, shear stresses between a physisorbed gold cluster and graphite depend only weakly on the sliding path.
Note that this should not be confused with those cases in which the angular misfit between the two sliding objects varies while the sliding orientation (associating with the wavelength of the substrate surface potential) does not.
In the latter scenario, the six-fold symmetry still holds.
It should also be discriminated from (quasi-)static calculations where atoms are in thermal equilibrium at any given moment, from which the static friction and the potential effect of contact aging can be deduced \cite{Li2011N,Feldmann2014PRL}, which are out of the scope of this study.
Although nominally rigidity holds the upper hand against elastic deformation and dislocations, sliding-induced vibration of individual atoms about their non-equilibrium steady-state positions annihilates the accumulation of instantaneous forces systematically, thereby a vanishing of mean lateral force (i.e., structural lubricity) arises.
For a finite-sized geometry, instability can be introduced from a propagating contact line normal to the sliding direction as a result of  the generically asymmetric structure of the slider \cite{Gao2022FC}, although such an effect will decay with cluster size in terms of the scaling argument \cite{Mueser2001PRL,Dietzel2013PRL}.

When the cluster is free, i.e., no constraints are applied except for the springs, non-monotonic dependency of shear stresses on $k_{\rm spr}$ is observed in the MD simulations, with a turning point located in the proximity of $k_{\rm spr}$=500 N/m, as shown  in Fig.~\ref{fig_sig_k}. 
Please note that this particular value itself may not be practically significant, but regardless, it brings qualitative discernment into the effect of driving spring stiffness on shear stresses in the structurally lubric regime from a computational point of view.
Beyond this critical value, shear stress decreases gradually with increasing driving spring stiffness, which appears to be in line with predictions from the Prandtl model in the underdamped regime ($\eta\gtrsim$ 1).
In this case, do the results presented in Fig.~\ref{fig_sig_k} mean that a `free' cluster implies a transition of the model driving spring system from an overdamped to an underdamped regime? 
This will be discussed in a bit.
Regardless, higher shear stresses compared to those obtained from the rigid model are observed for free clusters, indicating an increase in the amount of dissipated energy. Given that the rigid cluster can be simply viewed as a free cluster with extended dimensions, this observation obeys the scaling argument.
Below the transition value for the driving spring stiffness shown in Fig.~\ref{fig_sig_k}, shear stress decreases with decreasing driving spring stiffness, until becoming comparable to the rigid model.
Such a propensity is however counter-intuitive since weak springs are more compliant and thereby supposedly lead to higher levels of energy dissipation (due to increased elastic instability), in the absence of thermal activation.
Now, the question is why?

\subsection{Spontaneous cluster rotation}
\begin{figure}[ht]
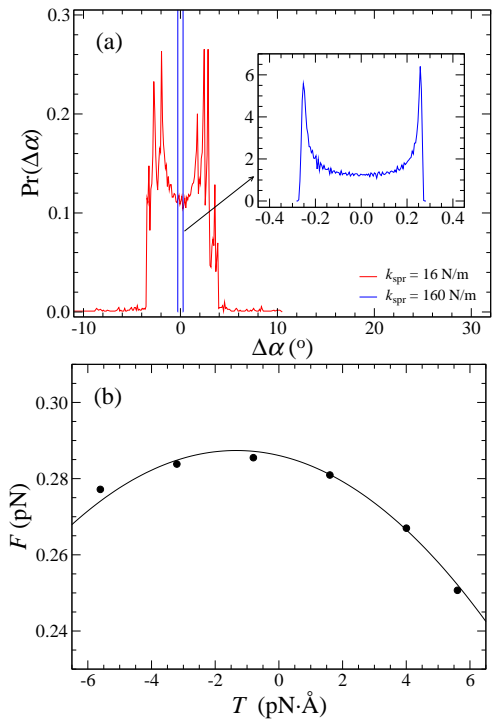

\centering
\includegraphics[width=0.35\textwidth]{angle_distribution.eps}
\includegraphics[width=0.355\textwidth]{torque.eps}~~
\caption{
(a) Probability of in-plane rotation angles $\Delta \alpha$ of a free cluster dragged by springs with stiffness values of $k_{\rm spr}$=16 (red) and 160 (blue) N/m. The inset shows a re-scaled view of the blue curve in (a). (b) A representative relationship between lateral force $F$ and instantaneous torque $T$ at $k_{\rm spr}$=16 N/m. Here, $T\propto|\Delta\alpha|$ given that the magnitude of $\Delta\alpha$ is always less than $\pi/6$.
}
\label{fig_ang_dist}
\end{figure}

From a fundamental point of view, individual atoms at a contact interface can be viewed as the smallest possible units that form a multi-asperity contact, which is analogous to what is depicted in a high-dimensional Frenkel-Kontorova model \cite{Frenkel1939}.
As long as the perturbation of each atom at the sliding interface is small enough (typically less than a unit lattice), the ensuing dissipation can fall into the description of linear-response theory \cite{Adelman1976JCP}.
Along this line, an interesting observation is that springs with high compliance promote oscillatory cluster rotation.
This is evidenced by the distribution of relative rotation angles ($\Delta\alpha$) of a sliding cluster, as shown in Fig.~\ref{fig_ang_dist}a, whose in-plane RDoF is highly inhibited when it is linked to a stiff spring as opposed to a compliant one.
It has been already known that even a small angular mismatch between the sliding surfaces can alter the interfacial commensurability \cite{Wijn2012PRB}. This results in the damping coefficient (and thus, the shear stress) being a rotation-angle-dependent function.
The lower lateral forces observed at a larger $|\Delta\alpha|$, as shown in Fig.~\ref{fig_ang_dist}b, can be attributed to spontaneous, rotation-induced depinning of the cluster subject to torques ($T$) that arise from the interfacial interaction, which is also in alignment with the Arrhenius equation such that the interplay may change under the effect of sliding speed \cite{Guerra2010NM}.
%

\begin{figure}[ht]
\centering
\includegraphics[width=0.45\textwidth]{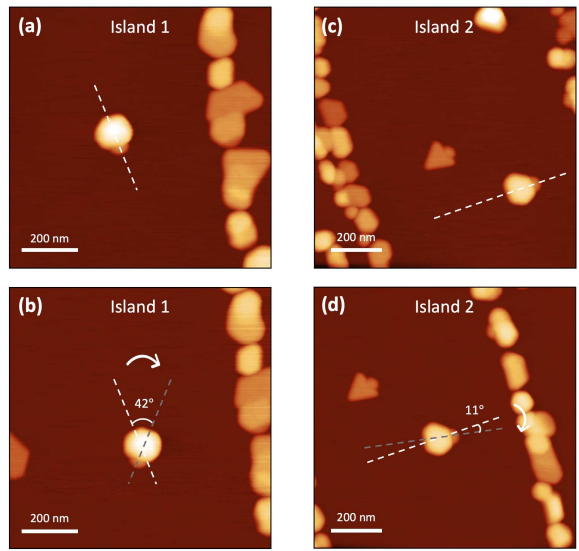}
\caption{
AFM topography images of two representative gold clusters (a, c) before and (b, d) after manipulation, showing spontaneous rotations of (b) 42\textdegree{} and (d) 11\textdegree{}, respectively. The experiments were conducted with an applied normal load of 0.0 nN, i.e., under adhesive interaction only.
}
\label{fig_rotation_afm}
\end{figure}

It is important to note that experiments involving `tip-on-top' manipulations of gold clusters on graphite often result in spontaneous rotations of the cluster (see Fig.~\ref{fig_rotation_afm} for two representative examples).
Specifically, for this structurally superlubric material system \cite{Cihan2016NC}, the static friction between the gold cluster and graphite substrate is significantly smaller than the static friction between the tip apex and the cluster.
Thus, when the AFM tip is placed on top of a cluster (hence the phrase `tip-on-top' \cite{Dietzel2009APL}) and scans a small area in contact mode, the tip will effectively drag the cluster to slide laterally on the substrate.
As described in Fig.~\ref{fig_rotation_afm}, spontaneous rotations of the cluster within the range of 0 – 75\textdegree{} are often observed during the manipulations.

The experimental observations of spontaneous rotations can be attributed to two main reasons:
(i) Experimentally, it is highly unlikely for the tip to be placed directly on top of the cluster’s in-plane COM. Thus, a torque will inevitably be introduced during the manipulation process and 
(ii) the adhesion force between the tip and the cluster is not strong (mainly due to small contact areas), such that the tip may occasionally slip during the reciprocal sliding action.
These slips may introduce torsional forces as well.
These experimental observations support the simulation results, where clusters are spontaneously rotated and depinned due to torques.
However, the range of in-plane rotation angles for the experiments are significantly larger compared to the simulations.
This result is expected because the experimental lateral stiffness of the tip-cluster contact $k_{\rm spr}$ is expected to be significantly smaller (on the order of only a few N/m \cite{Socoliuc2004PRL}) than the simulated lateral stiffness $k_{\rm spr}$ values of 16 N/m and 160 N/m.
It is important to note here that the experimental shear stresses are within the range of 0.27 – 20 kPa (overlapping with the range of simulated shear stresses from Fig.~\ref{fig_sig_k}.).

\begin{figure}[ht]
\centering
\includegraphics[width=0.48\textwidth]{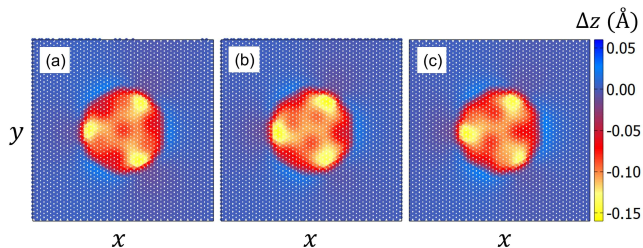}
\caption{Atomic normal displacement of the topmost-layer of the graphite substrate at three consecutive instances during sliding motion from (a) to (c). The relative shift of the cluster with respect to graphite in panels (b) and (c), with respect to panel (a), are 1.7 and 2.6 \AA, respectively. The in-plane positions of the cluster are centered for straightforward comparison}
\label{fig_moire}
\end{figure}

Theoretically, the dependence of friction on rotation angle can be rationalized in terms of the moir\'e patterns formed at the contact interface, which quantify the out-of-plane displacement (puckering) of the topmost graphite layer underneath the sliding cluster.
Characterization in terms of atom normal displacements rather than an in-plane description \cite{He1999Science} complies with the idea from Tomlinson \cite{Tomlinson1929} such that the discontinuous dynamics of the cluster are captured fairly well.
This is also in agreement with the nature of graphene as a 2D material, whereby the out-of-plane stiffness is much lower than the in-plane stiffness.
Structurally, the second-to-the-bottom atomic layer of a gold (111) cluster breaks the mirror symmetry generically, which would intensify interfacial atom vibration at the propagating contact line, especially when it is normal to the sliding direction \cite{Gao2022FC}.
Compared to a semi-infinite slab, instability of a circular cluster is located primarily at its circumference, where the highest contrast in the moir\'e patterns are visualized as shown in Fig.~\ref{fig_moire}.
This can be a reason why quasi-discontinuous motion of the cluster is not distinctively observed in the current simulations.
Besides, rotated Moir\'e patterns can effectively disturb the contact asperities from advancing in a collective manner.

\subsection{Energy dissipation via phonons}
\begin{figure}[ht]
\centering
\includegraphics[width=0.4\textwidth]{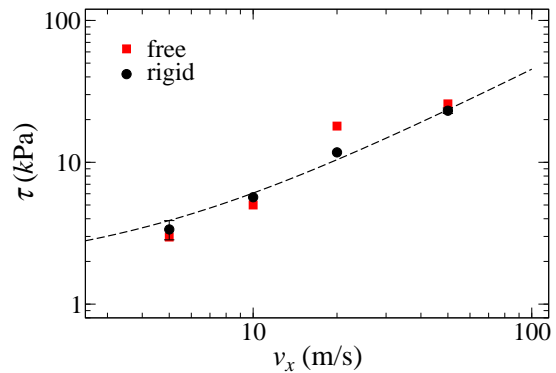}~~~~~~
\caption{Velocity-dependent shear stresses ($\tau$) of a free (red squares) and a rigid (black circles) cluster. The dashed line is a linear fit to the black data points. For both cases, $k_{\rm spr}$=1600 N/m.}
\label{fig_stress_vel}
\end{figure}

Friction describes an energy dissipation process, in which mechanical energy is converted into thermal energy via phonons as the main carrier.
In a non-equilibrium state, excessive phonons can be excited at the corresponding washboard frequency ($\nu_0$) \cite{Sokoloff1984SS, Braun2005PRE} such that more energy is dissipated via interfacial lattice vibration and thereby causes friction to rise \cite{Duan2021NL}.
Overall, Stokesian resistance results from the current model, as shown in Fig.~\ref{fig_stress_vel}, which manifests in quasi-elastic deformations of the contact bodies as well as inevitable hysteresis at the sliding interface.
However, outliers can also be found at some particular sliding speeds with high magnitudes that exceed typical statistical errors, e.g., the data point visualized by the red square at 20 m/s in Fig.~\ref{fig_stress_vel}.
To rationalize such an inconsistency, a vibrational spectrum $g(\nu)$ of atoms at the contact interface is deducted from MD-calculated instantaneous forces by means of fast Fourier transform (FFT) \cite{Heideman1985AHES}.
An equivalent result should also be obtained from the vibrational density of states converted from a normalized mass-weighted velocity auto-correlation function as demonstrated in an earlier simulation work \cite{Gao2021PRM}.

\begin{figure}[ht]
\centering
\includegraphics[width=0.49\textwidth]{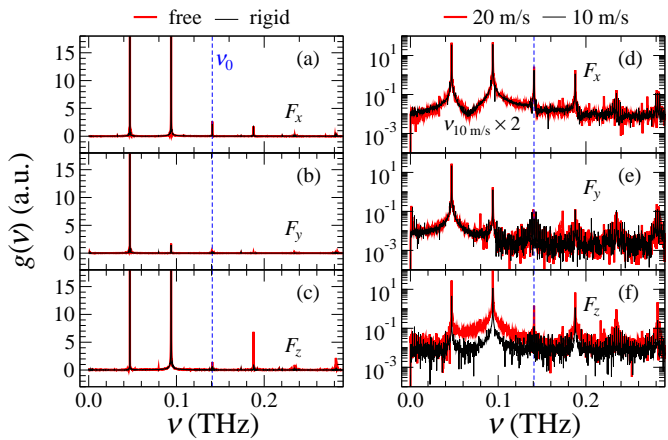}
\caption{Fourier transfer of the instantaneous lateral forces ($F_{x,y,z}$) for a free/rigid cluster sliding at 20 m/s (a-c) and a free cluster sliding at 10 or 20 m/s (d-f). The blue dashed lines represent the corresponding washboard frequency $\nu_0$=0.141 THz at 20 m/s. The frequencies from 10 m/s sliding are doubled to enable a direct comparison with those from 20 m/s sliding. For all cases, $k_{\rm spr}$=1600 N/m.}
\label{fig_frequency}
\end{figure}

In a phonon spectrum, the excited peaks, as shown in Fig.~\ref{fig_frequency}, indicate at what frequencies the force signals are correlated, or in other words, energy is dissipated.
Discrete modes arise for sliding speeds such as 10 and 20 m/s, with the corresponding major peaks located in the vicinity of the fractures of the washboard frequency, i.e., $\frac{1}{3}\nu_0$ and $\frac{2}{3}\nu_0$.
As indicated by the dashed lines in Fig.~\ref{fig_frequency}, the washboard frequency is calculated as $\nu_0=v/a$ \cite{Braun2005PRE}, where $a$ = 1.42 \AA~is the side length of a hexagonal graphite unit cell.
The correlation lengths converted from those peak frequencies, i.e., $2a$ and $3a$, reflect precisely the substrate lattice spacing, which depicts a repetitive underdamped single-slip \cite{Medyanik2006PRL} sliding motion of the cluster along the armchair direction of graphite.
It suggests that, although the sliding mode is `discontinuous' by definition, structural lubricity can be retained at a multi-asperity contact due to lattice vibration.
Resonance does not occur under such circumstances given the higher natural frequency of an isolated cluster $\nu_{\rm nat}=1/2\pi \sqrt{k_{\rm spr}/m}=0.244$ THz as well as an even higher resonant frequency of a sliding cluster of $\nu_{\rm res}=1/2\pi \sqrt{(k_{\rm spr}+k_{\rm con})/m}$, where $k_{\rm con}$ is the contact stiffness (an unknown that is much smaller than $k_{\rm spr}$) and $m$ the mass of the cluster.

While the phonon modes corresponding to the lateral forces ($F_x$) or the transverse forces ($F_y$) appear similar between the free (red) and the rigid (black) cluster, as shown in Fig.~\ref{fig_frequency}a and \ref{fig_frequency}b, the partially differing spectra of the normal forces ($F_z$) can be the origin of the friction discrepancy.
As shown in Fig.~\ref{fig_frequency}c, the missing frequency spike of a rigid cluster (black) at around 0.19 THz in contrast to a free cluster (red) indicates a lack of energy dissipated at this particular frequency.
Within this context, the overdamped normal forces in a rigid cluster may likely lead to a reduction of friction forces due to the partly suppressed atomic DoFs.
To this end, one can interestingly conclude that forces in the normal direction are inherently correlated with the lateral forces not only via their mathematical mean (Amontons' law) but also their vibrational frequencies.
Moreover, as shown in Fig.~\ref{fig_frequency}f, the spectrum intensity of the normal forces from a faster sliding cluster (red) is more pronounced than a slower one (black) across a certain range, which is an indication of enhanced energy dissipation in line with the Stokesian correlation between friction and sliding velocity.

\section{Conclusions}
In this work, MD simulations have been carried out to investigate kinetic friction between a crystalline gold cluster and a graphite substrate in the absence of contaminants.
The gold cluster, as a nominally rigid body, slides under dragging forces imposed by a virtual spring that models the tip-cluster interaction in an AFM-based manipulation experiment.
Beyond a critical value, shear stresses of a free cluster decrease with increasing spring stiffness, which is consistent with theoretical predictions from the Prandtl model in the underdamped regime ($\eta\gtrsim$ 1).
On the other hand, below the critical value, shear stresses decrease with decreasing spring stiffness. This is explained by spontaneous rotations of the cluster induced by elastic instability, which effectively reduces the energy barrier to sliding via a commensurability change; this is evidenced by an inverse friction-torque relationship.
Interestingly, such a non-monotonic dependency of shear stresses on driving spring stiffness is not observed when the translational DoFs of atoms in the cluster's topmost-layer are suppressed (thus resulting in a more `rigid' cluster or equivalently, a cluster with extended dimensions). In this case, shear stresses are independent of driving spring stiffness, with values less than one half of the maximum shear stress for a free cluster.
Contour of moir\'e patterns indicates that instability (in terms of normal displacements of atoms in the topmost-layer of graphite) at the circumference of a round cluster may not necessarily cause distinct discontinuous stick-slip motion. Instead, a Stokesian relationship between shear stress and sliding speed arises, whereby shear stresses increase linearly with increasing sliding speed.
Having said this, the asymmetric nature of the gold (111) surface parallel to the graphite basal plane can occasionally lead to the observation of high friction values in a free cluster that are beyond the range of statistical errors.
Approaching this observation by means of phonon vibrational spectra, one can find that, unlike free clusters, rigid clusters can suppress out-of-plane phonon excitation at certain frequencies, thus leading to suppressed shear stresses.

It is the hope of the authors that this work will bring new physical insights into friction variations observed during manipulation experiments performed on nanoscale clusters on atomically flat substrates, and also inspire new ideas with respect to friction control at small-scale, structurally lubric contacts.

\begin{acknowledgments}
H.G. thanks Martin H. M\"user for useful discussions. W.H.O. acknowledges support from the National Science Foundation (NSF) in the form of a Graduate Research Fellowship (GRF). 
This research was supported by the German Research Foundation (DFG) under grant number GA 3059/2-1 and the NSF under award number 2131976. 
\end{acknowledgments}

\appendix*

\section{Predictions from the Prandtl model}

Manipulation of a nanoscale cluster using an AFM tip \cite{Dietzel2007JAP, Cihan2016NC} can be idealized within the framework of the Prandtl model \cite{Prandtl1928ZAMM, Popov2012ZAMM} provided that the rigidity of the cluster exceeds the interfacial interaction strength such that the cluster can be treated as a point mass.
As a whole, the cluster slides over a 1D periodic potential landscape dragged by a virtual spring that models the tip-cluster interaction.
The associated effective stiffness ($k_{\rm eff}$) integrates three contributions in series, namely: (i) the torsional stiffness of the cantilever, (ii) the interaction strength between the tip apex and the cluster, and (iii) the interaction strength between the cluster and the substrate, in which the weakest one plays the predominant role \cite{Baykara2018APR}.
Here (i) and (ii) are combined as one assuming that the tip and the cluster are firmly attached.
In order to achieve ultra-low friction, $k_{\rm eff}$ should be large enough such that
${\eta}=4{\pi}^2V_0/k_{\rm eff}a^2$
($V_0$: energy barrier, and $a$: lattice spacing) is less than 1, at which sliding is smooth with extremely low energy dissipation \cite{Socoliuc2004PRL}.
If $k_{\rm eff}$ is too small, the point mass will be underdamped due to elastic instability and thereby single/multi-slip can occur under the influence of, e.g., normal load and thermal activation \cite{Sang2001PRL, Socoliuc2004PRL, Medyanik2006PRL}.

The Prandtl model is a reduced-order model that provides qualitative insights into small-scale friction, where the total potential energy of the system can be expressed as
\begin{equation}
V(x,t) =
V_0{\rm cos}(\frac{2\pi{x}}{a})-
\frac{1}{2}k_{\rm eff}(vt-x)^2,
\end{equation}
where $x$ denotes the position of the point mass, $v$ the speed of the dragging support, and $t$ the time.
Without thermal activation, the point mass starts to jump into the next potential well at a saddle point $V''(x,t)=0$, where the position of the support relative to the point mass determines the amount of energy to be dissipated.
Dynamics of the point mass can be described by a Langevin equation, written as
\begin{equation}
m\ddot{x}+m{\gamma}\dot{x}+k(x-vt)={\rm sin}x,
\end{equation}
where $m$ is the mass and $\gamma$ the damping coefficient.
To simplify the problem, parameters in this calculation are with reduced units with assigned values, i.e., $m=V_0=2\pi/a=1$ and $\gamma=0.1$, referring to a recent work by M\"user \cite{Mueser2020L} where more detailed explanations shall be found.
Kinetic friction $F_{\rm kin}$ reported in the following denotes a mean value of lateral forces over numerous stick-slip periods at steady-state.

\begin{figure}[ht]
\centering
\includegraphics[width=0.4\textwidth]{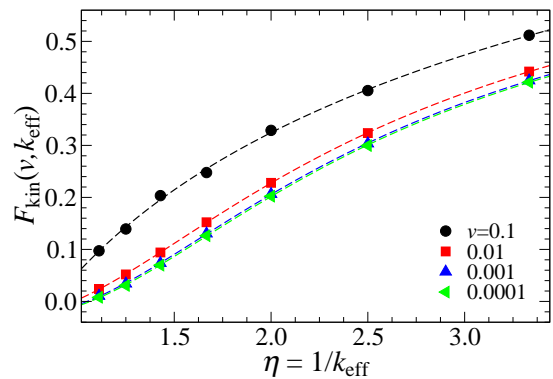}~~~~
\caption{Mean kinetic friction $F_{\rm kin}$ predicted from the Prandtl model as functions of $\eta$ (inversely related to the effective spring stiffness $k_{\rm eff}$) and sliding velocity $v$.}
\label{fig_prandtl}
\end{figure}

By substituting the chosen values into the expression of $\eta$, one can readily clarify the simple inverse relationship between $\eta$ and $k_{\rm eff}$, i.e., $\eta$ = 1/$k_{\rm eff}$.
Note, the $ad~hoc$ damping coefficient $\gamma=0.1$ is chosen to ensure that the sliding object is in an underdamped regime with single stick-slip motion being captured.
To this end, the more compliant the spring, the higher the elastic instability of the point mass and thereby more elastic energy is stored during the `stick' phase, which will be released subsequently while the object slips into the next potential well. 
As shown in Fig.~\ref{fig_prandtl}, kinetic friction increases with $\eta$ across a wide range of sliding speeds.
Given the current range of spring stiffnesses, a plateau that corresponds to a continuous sliding in an overdamped regime (i.e., when $\eta\leq$1) has not yet been reached.
In an athermal environment, resistance of the point mass in response to the dragging forces falls in the linear response regime \cite{Adelman1976JCP}, which leads to a linear dependence of kinetic friction on the sliding velocity, or in other words, Stokesian friction.
At low sliding velocities, the point mass is in thermal equilibrium at a given moment, beyond which a transition from a linear to a logarithmic dependence can take place when thermal effects kick in \cite{Mueser2020L}.

\nocite{*}

\bibliography{apssamp}

\end{document}